\documentclass[prl,twocolumn,showpacs,preprintnumbers,amsmath,amssymb,amsfonts]{revtex4}

\usepackage{graphicx}
\usepackage[dvips]{color}
\usepackage{dcolumn}
\usepackage{bm}
\usepackage{setspace}

\begin{document}

\title{Direct observation of charge order and orbital glass state in multiferroic LuFe$_2$O$_4$.}

\author{A.M. Mulders$^{1,2}$, S.M. Lawrence$^1$, U. Staub$^3$, M. Garcia-Fernandez$^3$, V. Scagnoli$^4$, C. Mazzoli$^4$, E. Pomjakushina$^{5,6}$,  K. Conder$^5$, Y. Wang$^7$.}
  \affiliation{$^1$Department of Imaging and Applied Physics, Curtin University of Technology, Perth, WA 6845,  Australia}
  \affiliation{$^2$The Bragg Institute, Australian Nuclear Science and Technology Organization, Lucas Heights, NSW 2234, Australia}
 \affiliation{$^3$Swiss Light Source, Paul Scherrer Institut, 5232 Villigen PSI,
Switzerland}
\affiliation{$^4$European Synchrotron Radiation Facility, BP 220, 38043 Grenoble Cedex 9, France}
\affiliation{$^5$Laboratory for Developments and Methods, Paul Scherrer Institut, 5232 Villigen PSI,
Switzerland}
\affiliation{$^6$Laboratory for Neutron Scattering, ETHZ \& PSI, 5232 Villigen, Switzerland}
\affiliation{$^7$Department of Applied Physics, The Hong Kong Polytechnic University, Hong Kong, China}

\date{\today}

\begin{abstract}
Geometrical frustration of the Fe ions in LuFe$_2$O$_4$ leads to intricate charge and magnetic order and a strong magnetoelectric coupling.
Using resonant x-ray diffraction at the Fe $K$ edge, the anomalous scattering factors of both Fe sites are deduced from the ($\textit{h}/{3}$ $\textit{k}/{3}$ $\textit{l}/{2}$) reflections.
The chemical shift between the two types of Fe ions equals 4.0(1) eV corresponding to full charge separation into Fe$^{2+}$ and Fe$^{3+}$. 
Polarization and azimuthal angle dependence of the superlattice reflections demonstrates the absence of differences in anisotropic scattering revealing random orientations of the Fe$^{2+}$ orbitals characteristic of an orbital glass state.
\end{abstract}


\maketitle
New materials that exhibit strong magnetoelectric coupling are fascinating because a large coupling between ferroelectric and magnetic interactions is rare, and its origin often unclear. 
Competing interactions lead to novel ground states that give rise to unusual material properties, i.e coexistence of spontaneous magnetic and ferroelectric order \cite{kimura_nature_2003}. The ability to control electric polarization with a magnetic field or the magnetization with an electric field 
\cite{zhao_natmat_2006, yamasaki_prl_2007, bodenthin_prl_2008} 
makes these multiferroic materials promising candidates for novel applications such as 4-state memory and switchable magneto-optical devices. In most ferroelectric materials, electric polarization arises from covalent bonding between anions and cations or the orbital hybridization of electrons. Alternatively ferroelectric polarization may arise from frustrated charge order as reported for LuFe$_2$O$_4$ \cite{ikeda_nature_2005}. This compound is of particular interest because, in addition to ferroelectricity, magnetism originates from the same Fe ions and this holds the promise of strong magnetoelectric coupling. The ferroelectric and magnetic order take place at and near ambient temperature which provides the potential for room temperature multiferroics.

The crystal structure of LuFe$_2$O$_4$ consists of a triangular double layer of iron ions, forming trigonal bipyramids with five oxygen nearest neighbors, in which an equal amount of Fe$^{2+}$ and Fe$^{3+}$ are believed to coexist at the same site \cite{sekine_jssc_1976}. The occurrence of different charge order schemes has already been studied in more detail theoretically and experimentally \cite{xiang_prl_2007, zhang_prl_2007}.
LuFe$_2$O$_4$ adopts a ferroelectric ground state below $\sim$350 K, while below $\sim$250 K  two-dimensional magnetic order is established in the triangular planes which enhances the ferroelectric polarization by 20\%, illustrating coupling between the magnetic and ferroelectric order \cite{ikeda_nature_2005,iida_jpsj_1993}. 
The observation of the (1/3 1/3 13/2) reflection using resonant x-ray Bragg diffraction (RXD) further supports the existence of charge ordering \cite{ikeda_jpcm_2008}.

RXD has become a powerful technique to study charge, orbital and magnetic arrangements. Tuning the energy of the incoming radiation to an absorption edge permits recording Bragg reflections with enhanced sensitivity to the specific ion and its electronic configuration. 
In case of the Fe $K$ edge, the incident x-rays virtually excite an electron from the 1s core level to the empty 4p states, followed by a decay of the  electron back to the core hole. 
This effect results in a significant variation in the atomic scattering factors of the Fe ions for x-ray energies close to the Fe $K$ edge.
The atomic scattering factors are also affected by variations in charge state, ordered aspherical electron densities or ordered magnetic moments. In particular asphericity of the atomic electron density results in anisotropy of the tensor of x-ray susceptibility (ATS).
Each of these phenomena has a specific dependence on the polarization of the incoming radiation and the orientation of the sample with respect to the scattering geometry.
The significance of RXD has been demonstrated, among others, in the manganites \cite{herrero_prb_2004}, nickelates \cite{scagnoli_prb_2005} and magnetite \cite{nazarenko_prl_2006}.

In a perfectly charge ordered state, each site may be considered as having an excess and a deficiency of half an electron respectively, compared to the average ion valence of Fe$^{2.5+}$.  Alternatively charge disproportionation with fractional charges may exist as exemplified in nickelates \cite{alonso_prl_1999,staub_prl_2002}. 
Frustration arises because every excess charge prefers a deficiency charge as a neighbor which is not possible on a triangular lattice.
However, in the presence of a second triangular layer a net transfer of charge from the first layer to second occurs because then it is possible to have charge order on each layer in a honeycomb lattice arrangement \cite{wannier_pr_1950, nasu_prb_2008}.

The crystal field of the trigonal bipyramids splits the 3$d$ states of LuFe$_2$O$_4$ into two doublets ($d_{xy}/d_{x^2-y^2}$ and $d_{xz}/d_{yz}$) and a singlet ($d_{z^2}$) \cite{nagano_prl_2007}.
Fe$^{3+}$ with five 3d electrons is spherical while Fe$^{2+}$ with six 3d electrons exhibits doubly degenerate orbital degree of freedom in the $d_{xy}/d_{x^2-y^2}$ ground state.

In this paper we present RXD data with azimuthal angle and polarization analysis and our results clarify that the charge order is close to electronic states of Fe$^{2+}$ and Fe$^{3+}$, in contrast to small values of disproportionation observed in nickelates  \cite{staub_prl_2002} and manganites \cite{herrero_prb_2004}.
Moreover the absence of scattering due to ATS demonstrates
a glass state of the Fe$^{2+}$ orbitals in agreement with calculations \cite{nagano_prl_2007}.

Polycrystalline LuFe$_2$O$_4$ was prepared by a solid state reaction as reported in ref. \cite{iida_jcg_1990}. Starting materials of Lu$_{2}$O$_{3}$ and Fe$_{2}$O$_{3}$ with 99.99\% purity were mixed, pressed into pellets and sintered at 1200$^\circ$C during 6 h in H$_2$/He/CO$_2$ atmosphere (H$_2$/CO$_2$ ratio 1/3) and quenched into ice water. After grinding, the obtained powder was hydrostatically pressed and sintered at the same conditions during 3h.
The crystal growth was carried out using Optical Floating Zone Furnace with four 1000W halogen lamps as a heat source, growth rate 1 mm/h, 2 bar pressure of CO­$_2$/CO mixture (5/2 ratio).
The single crystal was cut and samples have been polished perpendicular to the [001] and [110] directions. The magnetic ordering temperature was determined with a SQUID magnetometer at 240 K and pyroelectric current measurement confirmed ferroelectric order below $\sim$330~K and enhanced ferroelectric order below $\sim$220 K. 

Various ($\textit{h}/{3}$ $\textit{k}/{3}$ $\textit{l}/{2}$) superlattice reflections were recorded at the Fe $K$ edge at beamline ID20 of the ESRF \cite{paolasini_jsr_2007}. Polarization analysis was performed using a MgO (2 2 2) analyser crystal of which the polarization efficiency was determined at 0.98 for the energy of the Fe $K$ edge. 
In addition, RXD was recorded at the MS beamline at the SLS (see also supplement \cite{supplement}) using the Pilatus 2D detector  \cite{broennimann_jsr_2006}.
The background, mainly originating from the fluorescence of the sample, was determined from selected border regions of the area detector and subtracted. The integrated diffracted intensity was corrected for polarization, absorption and sample geometry.  
The absorption, $\mu(E)$, was obtained from the (006) reflection by iteration of the calculated anomalous intensity according to spacegroup $R\bar{3}m$ and $\mu(E)$ deduced from the ratio between calculated and integrated intensity recorded with the 2D detector.
Besides $\mu(E)$ was recorded at the X-ray Absorption Spectroscopy beamline at the Australian Synchrotron using powdered LuFe$_2$O$_4$ pressed with cellulose and Fe foil as energy calibration. It confirmed the validity of the method to obtain $\mu(E)$ from the (006).
RXD recorded at the SLS was used to analyse the energy dependence while RXD recorded at ID20 was used to investigate polarization dependence and rotation about the scattering vector {\bf q} (azimuthal angle $\Psi$). $\Psi$ is defined zero when [110] and [001] are in the scattering plane. 

\begin{figure}[t]
\vspace{3cm}
\begin{center}
\includegraphics[scale=0.35,bb = 540 70 41 475]{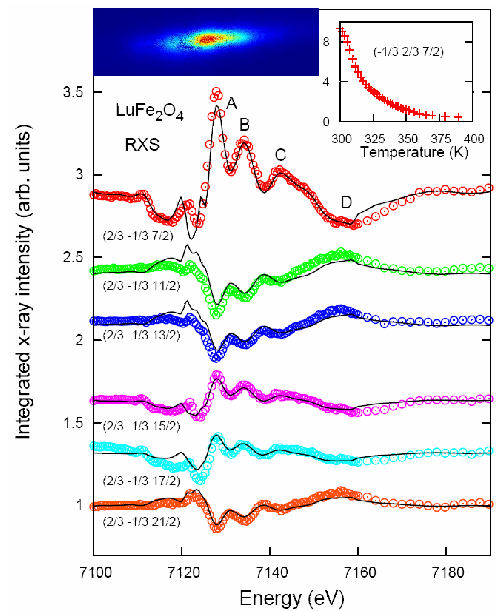}
\end{center}
\vspace{-.5cm}
\caption{ 
Integrated resonant diffraction intensity of the charge order (2/3 -1/3 $\textit{l}$/2) reflections recorded with 2D detector at T=15 K (see inset top left for $\textit{l}$ = 11), corrected for background, polarization, absorption and sample geometry, normalized (shifted for clarity) and compared to charge order model described in the text.
The inset (top right) shows that the superlattice intensity gradually disappears at the ferroelectric ordering temperature. For A, B, C and D see text.
}
\label{fig1}
\vspace{-.2cm}
\end{figure}

The ($\textit{h}$/3 $\textit{k}$/3 7/2) reflections show identical RXD \cite{supplement} while the magnitude and sign of the RXD observed at the (2/3 -1/3 $\textit{l}$/2) reflections depends on $\textit{l}$ (Fig. \ref{fig1}).
The relative magnitude and shape is constant between 10~K and 300~K and the diffracted intensity gradually disappears above the ferroelectric ordering temperature as illustrated in the inset of Fig. \ref{fig1}. 
Thomson scattering associated with the crystallographic distortion that accompanies the ferroelectric polarization dominates the diffracted intensity before and after the edge.
The variation of XRD with $\textit{l}$ suggests that the scattering amplitude related to the anomalous diffraction of the Fe ions adds phase shifted contributions to the scattering amplitude from the structural distortion, depending on the $\textit{l}$ index.
The structure factor is $F = \sum_j f_j \exp (i \textbf{q} \cdot \textbf{r}_j)$ with $f_j$ the atomic form factor of atom $j$ and $\textbf{r}_j$ its position in the unit cell.
To analyse the energy dependent intensity we separate $F$ into an energy independent and an energy dependent term, $F = F^0_{Fe,Lu,O} + F(E)_{Fe}$, where the first term is the Thomson scattering of Fe, Lu and/or O ions and the second term is anomalous diffraction due to the Fe ions. 
These terms are written as
$F^0_{Fe,Lu,O}=A + iB$ and $F_{Fe}(E) = \sum _j (f_j'(E) + if_j''(E)) e^{i q \cdot r_j}$ where
$A$ and $B$ are the real and imaginary component of the non resonant structure factor and $f_{j}'(E)$ and $f_{j}''(E)$ are the real and imaginary component of the anomalous scattering factor of the Fe ions. $f_{j}'(E)$ and $f_{j}''(E)$ are related through the Kramers-Kronig (KK) relation.

\begin{figure}[t]
\vspace{5.7cm}
\begin{center}
\includegraphics[scale=0.45,angle=-90, bb = 550 60 130 490]{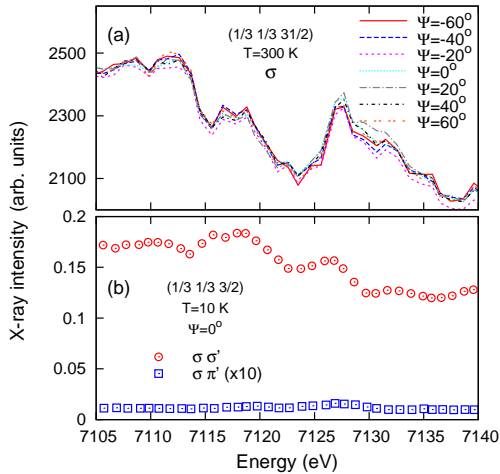}
\end{center}
\vspace{-.7cm}
\caption{ (a) Intensity of the (1/3 1/3 31/2) reflection as function of azimuthal angle $\Psi$. 
(b) Intensity of the (1/3 1/3 3/2) reflection recorded with $\sigma \sigma'$ and $\sigma \pi'$ radiation. The latter intensity is multiplied by 10 for clarity.
}
\label{fig2}
\vspace{-.5cm}
\end{figure}

For the superlattice reflections of this study $F_{Fe}(E)$ almost cancels except for the fact that Fe ions in different local electronic environments exhibit different energies of the $1s$ and the $4p$ states. This results in different transition energies, for example a chemical shift of 4.5 eV has been reported for $f_{j}''(E)$ between Fe$^{2+}$ and Fe$^{3+}$ in water complexes \cite{benfatto_cp_2002}.
In addition, ATS arises when the $4p$ states are split. 
The extended $4p$ orbitals are sensitive to local distortions and orbital order of the $3d$ shell gives rise to splitting of the $4p$ states, for example via the associated Jahn Teller effect. 

Differences in ATS give rise to a modulation in RXD intensity as function of azimuthal angle that is related to the symmetry of the distortion. Combined with the Thomson scattering, whose diffracted intensity is independent of azimuthal angle, this results in a change of the relative amplitude. Fig.\ref{fig2}a shows this is not observed in LuFe$_2$O$_4$ in contrast to orbitally ordered manganites \cite{herrero_prb_2004}. Furthermore, a significant contribution in diffraction with rotated polarization is expected for ATS, whereas polarization analysis shows that the diffracted intensity with rotated polarization, $\sigma \pi^\prime$, is weak and accounted for by the unrotated $\sigma \sigma^\prime$ contribution within the polarization resolution of the analyzer (See Fig. \ref{fig2}b).
The (1/3 1/3 $\textit{l}$/2) reflections of Fig. \ref{fig2} have arbitrary angles $\phi_\textit{l}$ with the [001] direction ($\phi_{3}$ = 73$^\circ$ and $\phi_{31}$ = 18$^\circ$) so that ATS is not canceled to zero by symmetry if there is orbital order. Moreover, polarization analysis of various other (1/3 1/3 $\textit{l}$/2) reflections and several azimuthal angles recorded for $\textit{l}$=29 yielded the same result \cite{supplement}.

Both aspects signal the absence of local asymmetric distortions associated with orbital order and
point to a rather symmetrical expansion or contraction of the trigonal bipyramids in the $ab$ plane, changing the electronic density at the resonant ion while the symmetry of the scattering factor is unaltered. 

\begin{figure}[t]
\vspace{5.7cm}
\begin{center}
\includegraphics[scale=0.45,angle=-90, bb = 550 70 150 475]{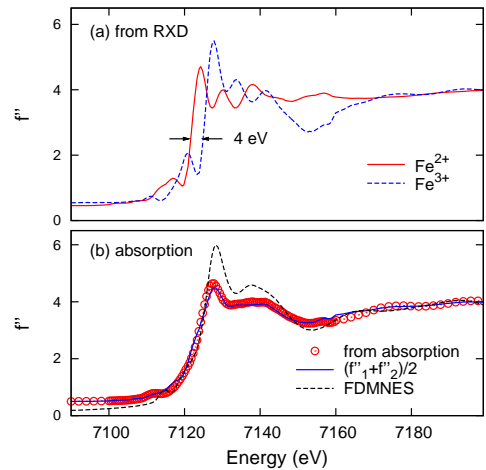}
\end{center}
\vspace{-.7cm}
\caption{ (a) $f_1''(E)$ and $f_2''(E)$ deduced from the refinement of RXD (Fig. \ref{fig1}) via the method described in the text and (b) its average compared with FDMNES calculation in $R\bar{3}m$ (dashed curve) and $(f_1''(E)+f_2''(E))/2$ from $\mu(E)$ (open circles).
}
\label{fig3}
\vspace{-.5cm}
\end{figure}

\begin{figure}[t]
\vspace{4.3cm}
\begin{center}
\includegraphics[scale=0.45,angle=-90, bb = 500 60 200 465]{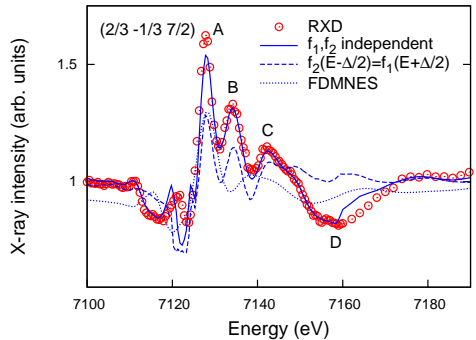}
\end{center}
\vspace{-.8cm}
\caption{ RXD compared to three models of charge order as described in the text. The solid curve corresponds to $f_{1,2}''(E)$ as presented in Fig. \ref{fig3}. The dashed and dotted curves correspond to identical $f_{1}(E)$ and $f_{2}(E)$ except for a chemical shift. The dashed curve is fitted to the RXD data whereas the dotted corresponds to $f_{1,2}(E)$ calculated with FDMNES.
}
\label{fig4}
\vspace{-.5cm}
\end{figure}

The energy dependencies of the ($\textit{h}$/3 $\textit{k}$/3 $\textit{l}$/2) reflections are examined with two different charge states of the Fe ions, labeled 1 and 2, and the structure factor at ($\textit{h}$/3 $\textit{k}$/3 $\textit{l}$/2) equals
\begin{eqnarray}
F_l(E) &\propto& A'_l + iB'_l \nonumber\\ 
&&+ f_1'(E) + if_1''(E) - f_2'(E) - if_2''(E).
\label{struct_fact}
\label{eq2}
\end{eqnarray}
Where $A'_l$ and $B'_l$ are real constants of arbitrary magnitude. In particular $B'_l$ is non zero due to the charge order and breaking of inversion symmetry.
The sum $(f''_1 + f''_2)$ is obtained from $\mu(E)$ using the optical theorem.
In this work we aim to deduce $f_1(E)$ and $f_2(E)$ from the RXD spectra  without any assumptions on the local distortion. To test the robustness of the specific energy dependencies, the series of (2/3 -1/3 $\textit{l}$/2) reflections are refined with different methods. Besides fully independent $f_1(E)$ and $f_2(E)$, we added the constraint $f_2(E-\frac{1}{2}\Delta) = f_1(E+\frac{1}{2}\Delta)$, where $\Delta$ equals the chemical shift. Moreover, we have calculated $f''(E)$ with FDMNES using the muffin tin approximation \cite{joly_prb_2001, supplement} in $R\bar{3}m$ and used $f_2(E-\frac{1}{2}\Delta) = f_1(E+\frac{1}{2}\Delta)$ to fit the XRD data. The chemical shift between $f_1''(E)$ and $f_2''(E)$ is similar in the three refinements and equals 4.0(1) eV. 
The first refinement resulted in distinct energy dependencies for $f_1(E)$ and $f_2(E)$. This is understood as a result of the experimental uncertainty of the XRD spectra and minimizing $f_2(E-\frac{1}{2}\Delta)- f_1(E+\frac{1}{2}\Delta)$ was added in a further refinement to promote similarity between $f_1(E)$ and $f_2(E)$.
Fig.\ref{fig1} illustrates the resulting fits and $f''_1(E)$ and $f''_2(E)$ are presented in Fig. \ref{fig3}. 
Both $f''_1(E)$ and $f''_2(E)$ exhibit characteristic features at similar energies above the edge however their magnitudes are different. Fig.\ref{fig4} compares the three different models for the (2/3 -1/3 7/2) reflection.  

The chemical shift between $f_1''(E)$ and $f_2''(E)$ of 4.0(1) eV corresponds to the chemical shift between Fe$^{2+}$ and Fe$^{3+}$ in FeO and Fe$_2$O$_3$ and confirms Fe$^{2+}$/Fe$^{3+}$ charge order. Besides the invariable chemical shift there are dissimilarities between the models. 
The double feature indicated with B,C in Fig. \ref{fig4} is not reproduced by the fit based on the FDMNES calculations. 
In addition, the broad feature, labeled D, cannot be accounted for with $f_2(E-\frac{1}{2}\Delta) = f_1(E+\frac{1}{2}\Delta)$. Yet, features B,C and D are either in or out of phase with the Thomson scattering (see Fig. \ref{fig1}). This signifies that the Fe-O bonds are distinct at each Fe site.

Long range of the Fe$^{2+}$ orbitals leads to ATS that is distinct from that of the spherical Fe$^{3+}$ ion.
Such difference in ATS between the Fe$^{2+}$ and Fe$^{3+}$ ions is absent and evidence of random orientations of the Fe$^{2+}$ orbitals. Our findings are consistent with the unconventional orbital state calculated by Nagano with frustrated orbital orientations and large degeneracy in the ground state \cite{nagano_prl_2007}.

The orbital state critically influences the superexchange interaction and is essential to understand the magnetoelectric properties of LuFe$_2$O$_4$.
In contrast to the manganites, where the orbital liquid state is discussed to be associated with the ferromagnetic and metallic state \cite{tokura_science_2000}, LuFe$_2$O$_4$ shows orbital disorder in the ferroelectric state. As such it is more appropriate to classify LuFe$_2$O$_4$ as an orbital glass,
adding further frustration to the already frustrated magnetic interactions of Ising spins on a triangular lattice.

Fe$_3$O$_4$ and LuFe$_2$O$_4$ both exhibit ferrimagnetic order combined with magnetoelectric effects \cite{khomskii_jmmm_2006} but their charge and orbital states are evidently distinct. Fe$_3$O$_4$
shows merely fractional charge order \cite{nazarenko_prl_2006} and recently order of the $t_{2g}$ orbitals was observed below the Verwey transition \cite{schlappa_prl_2008}.
While the superexchange is prevailing in Fe$_3$O$_4$, coulomb interactions dominate in
LuFe$_2$O$_4$, resulting in charge order and crystallographic distortions that do not accommodate orbital alignment. The reduced amplitude of $f''$ of Fe$^{2+}$ just above the edge (see Fig.\ref{fig3}a) is possibly indicative of this frustration and subject of further investigation.

Finally it is noted that cooling from 380 K in an electric field of $\sim$1 MV/m (EFC) did not result in a significant change of the ($\textit{h}/{3}$ $\textit{k}/{3}$ $\textit{l}/{2}$) superlattice reflections which contradicts the suggestion that ferroelectric charge order is the ground state only after EFC \cite{angst_prl_2008}.

In conclusion, our RXD data show an almost complete Fe$^{2+}$/Fe$^{3+}$ charge order as origin of the 
superlattice in LuFe$_2$O$_4$. The Fe scattering factors are isotropic and in agreement with frustrated and random orientations of the Fe$^{2+}$ orbitals, forming an orbital glass state.

We thank the beamline staff of X04SA and in particular Phil Wilmot and Dominique Meister for its excellent support, Chris Glover and the Australian Synchrotron for XANES measurement, Alec Duncan and the Centre for Marine Science and Technology 
for the use of Matlab routines, and acknowledge the ESRF for provision of beamtime. This work was partly performed at the SLS of the Paul Scherrer Institut, Villigen, Switzerland. 
We acknowledge financial support from the NCCR MaNEP Project, the Swiss National Science Foundation, AINSE and Access to Major Research Facilities Programme which is a component of the International Science Linkages Programme established under the Australian Government's innovation statement, Backing Australia's Ability.

\newpage

 \begin{center}
\large{\bf Supplement}\\
 \end{center}

\noindent In this supplement we present additional details of our study that might be of interest for the specialist reader. In the first section we present details of the FDMNES calculations, in the second section details of resonant x-ray diffraction (RXD) recorded at other superlattice reflections and in the third section details of the experimental geometry.

\section{1. FDMNES}
\noindent 
The FDMNES calculation presented in Fig 3.a of the paper was performed using the muffin tin approximation \cite{joly_prb_2001s} with space group $R\bar{3}m$ (no. 166), crystal parameters $a$ = 3.4406, $b$ = 3.4406, $c$ = 25.280, $\alpha$ = 90$^\circ$, $\beta$ = 90$^\circ$, $\gamma$ = 120$^\circ$ and atom positions of Lu at (0,0,0), Fe at (0, 0, 0.21518) and O at (0, 0, 0.1291) and (0, 0, 0.2926). The result converged at a cluster radius of 6$\AA$.

\section{2. In plane symmetry}
\noindent The symmetry of the ferroelectric structure of LuFe$_2$O$_4$ is not unambiguously known at this stage and there are possibly several tensors of x-ray susceptibility (ATS) allowed. These terms are significant in systems where charge order is connected with orbital order, such as manganites and magnetite, resulting in strong azimuthal angle dependences as well as observable contributions in the rotated channels \cite{grenier_prb_2004s, herrero_prb_2004s}. In contrast, charge order without orbital order as observed in for example NdNiO$_3$, results in RXD that is independent of azimuthal angle \cite{scagnoli_prb_2005s}.
The (1/3 1/3 31/2) and (1/3 1/3 3/2) reflections have angles $\phi$ = 18$^\circ$ and 73$^\circ$ with the [001] direction so that ATS is not canceled to zero by symmetry if there is orbital order. 
Moreover we recorded azimuthal angles of 0$^\circ$, 30$^\circ$ and 60$^\circ$ for the (1/3 1/3 29/2) reflection ($\phi$ = 19$^\circ$) and polarization dependencies for various (1/3 1/3 $\textit{l}$/2) reflections with $\textit{l}$=23-37 ($\phi$ = 23$^\circ$-15$^\circ$) (at beamline ID20 of the ESRF) which yielded the same result. 

Figure \ref{fig1} illustrates resonant X-ray diffraction (RXD) for ($\textit{h}$/3 $\textit{k}$/3 7/2) reflections ($\phi$ = 54$^\circ$) recorded at the MS beamline of the SLS. The RXD of these six reflections is identical, consistent with eq. (1) in the paper.

\begin{figure}[t]
\vspace{6.5cm}
\begin{center}
\includegraphics[scale=0.58,bb = 550 60 240 480, angle=-90]{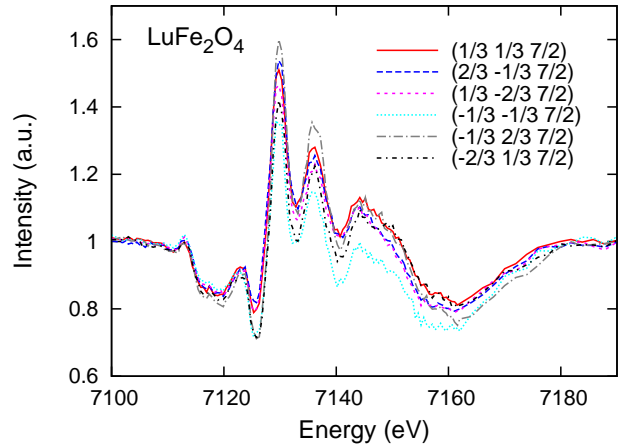}
\end{center}
\vspace{-.5cm}
\caption{
Normalized integrated resonant diffraction intensity of the charge order ($\textit{h}$/3 $\textit{k}$/3 7/2) reflections recorded with 2D detector at T=15 K, corrected for background, polarization and absorption. 
}
\label{fig1}
\vspace{-.2cm}
\end{figure}

\begin{table}[hb]
\caption{\label{table}Orientation of the sample ($\alpha$, $\omega_v$), detector ($\gamma$, $\delta$) and the polarization of the incident radiation ${\bf P}_\zeta$ for selected superlattice reflections recorded at the surface diffraction endstation of beamline X04SA with incident x-ray energy 7112.4~eV.
}
\begin{tabular}{@{}crrrrr}
\hline \hline
($\textit{h}$,$\textit{k}$,$\textit{l}$)& $\alpha$ & $\omega_v$ & $\gamma$ & $\delta$ &{\bf P}$_\zeta$\\
\hline
\\
(1/3, 1/3, 7/2)&~~10.00~&~~~66.29~&~~14.09~&~~19.73~&~~0.37\\
(1/3, -2/3, 7/2)&10.00&-48.32&12.97&20.31 &0.46\\
(-1/3, -1/3, 7/2)&10.00&-106.19&14.39&19.29&0.33\\
(-1/3, 2/3, 7/2)&10.00&127.84&15.70&18.44&0.21\\
(-2/3, -1/3, 7/2)&10.00&-169.65&15.48&18.27&0.21\\
(2/3, -1/3, 7/2)&10.00&10.06&12.69&20.49&0.49\\
(2/3, -1/3, 11/2)&10.00&11.39&21.03&20.99&0.07\\
(2/3, -1/3, 13/2)&10.00&14.56&25.27&21.05&-0.10\\
(2/3, -1/3, 15/2)&10.00&17.15&29.75&21.43&-0.23\\
(2/3, -1/3, 17/2)&10.00&20.69&34.61&20.83&-0.38\\
(2/3, -1/3, 21/2)&10.00&30.45&43.68&20.01&-0.56\\

\hline \hline
\end{tabular}
\end{table}

\section{3. Surface Diffractometer}
\noindent RXD as illustrated in Fig. 1 of the paper and Fig. \ref{fig1} of this supplement were
recorded at the surface diffractometer of the X04SA beamline \cite{website} using the Pilatus 2D detector \cite{broennimann_jsr_2006s}. This station is a large 2+3-circle surface diffractometer with two circles for the sample and three for the detector, plus a hexapod for precise alignment of the sample surface relative to the diffractometer axes. The diffractometer was operated with a vertical sample surface orientation.

Horizontal polarization of the incoming radiation was used which corresponds to $\pi$ incident polarization when the scattering plane is horizontal. When the scattering plane is not horizontal the incident beam contains both $\sigma$ and $\pi$ polarized components. The polarization of the incident radiation is most conveniently characterized by the Poincar\'{e} vector ${\bf P}_\zeta$ which equals +1(-1) for $\sigma$ ($\pi$) incident radiation \cite{blume_prb_1988s}. Table \ref{table} gives ${\bf P}_\zeta$ for the superlattice reflections presented in this study.
Table \ref{table} also summarizes the angles that define the orientation of the sample ($\alpha$, $\omega_v$) and detector ($\gamma$, $\delta$) with respect to the incoming radiation. The sample is mounted at the centre of rotation and $\alpha$ and $\gamma$ are the angles in the horizontal plane between the incoming beam and the sample surface and detector position respectively. $\omega_v$ is the sample rotation about an axis normal to the sample surface and $\delta$ is the angle between the horizontal plane and the detector position respectively.
No polarization analysis of the diffracted beam was performed and the pixel detector records both rotated and unrotated polarizations.

\end{document}